\documentclass[twocolumn,aps,showpacs,prb,amsmath,amssymb,floatfix]{revtex4-1} %final
\usepackage{longtable}
\usepackage{textcomp}
\usepackage{graphicx}% Include figure files
\usepackage{dcolumn}% Align table columns on decimal point
\usepackage{bm}% bold math
\usepackage{rotating}
\usepackage{color}
\def\ket#1{|\,#1\,\rangle}                 % ket
\def\bra#1{\langle\,#1\,|}                 % bra
\def\skp#1#2{\langle\,#1\,|\,#2\,\rangle}  % scalar product
\usepackage{amsmath}

\begin{document}

\preprint{submitted \date{\today}}
\title{Many-body theory of ultrafast demagnetization and angular momentum transfer
in ferromagnetic transition metals}

\author{W. T{\"o}ws}
\author{G. M. Pastor}
\affiliation{
Institut f{\"u}r Theoretische Physik,
Universit{\"a}t Kassel,
Heinrich-Plett-Stra{\ss}e 40, 34132 Kassel, Germany
            }

\begin{abstract}

Exact calculated time evolutions in the framework of a many-electron
model of itinerant magnetism provide new insights into the 
laser-induced ultrafast demagnetization observed in ferromagnetic (FM) 
transition metal thin films. The interplay between local spin-orbit 
interactions and interatomic hopping is shown to be at the origin of 
the observed post-excitation breakdown of FM correlations between 
highly stable local magnetic moments. 
The mechanism behind spin- and angular-momentum transfer is revealed 
from a microscopic perspective by rigorously complying with all
fundamental conservation laws. An energy-resolved analysis of the 
time evolution shows that the efficiency of the demagnetization process 
reaches almost 100\% in the excited states.

\end{abstract}

\date{\today}

% Classification Scheme.
%\keywords{Suggested keywords}%Use show-keys class option if keyword
%                              %display desired
%
%\vspace{2mm}

\maketitle

Pump-and-probe femtosecond-laser experiments on thin magnetic transition metal (TM) films have 
shown, almost 2 decades ago, that the ferromagnetic (FM) order breaks 
down on a time scale of a few hundred femtoseconds after the pulse 
absorption.\cite{Bea96} This remarkable finding opened the way to 
an ever since growing research field, which has a wide fundamental and 
practical importance.\cite{Hoh97,Koo00,Rhi03,Lis05,Cin06,Sta07,Stan07,Car08,Sta10,Rad11,Ber14} 
The phenomenon as such can be neither an immediate consequence of the 
excitation, since optical transitions conserve spin, nor the result of thermally activated stochastic processes, which 
would involve a much longer time scale. Instead, ultrafast demagnetization (UFD) reflects the 
intrinsic many-body dynamics of correlated excited electrons in FM metals. Understanding its 
origin and controlling its properties is therefore of crucial
importance.\cite{Kir10}

From a fundamental perspective, it has been clear from the start that 
spin-orbit (SO) interactions must play a central role in the dynamics,
since only the relativistic corrections break the conservation of the 
total electronic spin.\cite{Rhi03,Zha00,Big02} Indeed, the spin-orbit coupling (SOC) allows exchanges 
between the dominant local $3d$ spin moments $\vec s_i$ and the local 
orbital moments $\vec l_i$ at every TM atom $i$. Taking into account 
that the total angular momentum $\vec j_i = \vec l_i + \vec s_i$ is 
conserved in this process, the focus of attention quickly moved towards 
quantifying the time dependences of the spin and orbital moments, and to correlate them to the observed demagnetization.
This has been experimentally achieved by performing time-resolved X-ray magnetic circular dichroism (XMCD) measurements 
on Ni.\cite{Sta07,Sta10,Boe10} These works showed, in contrast to early 
expectations, that no enhancement of $l_{iz}$ accompanies the decrease of
$s_{iz}$, but rather that both $s_{iz}$ and $l_{iz}$ decrease as a function 
of time with a time constant of $\tau\simeq 120$~fs. Since $\vec l_i$ is not a 
reservoir for angular momentum, the authors concluded that a femtosecond spin-lattice
relaxation, i.e., a substantial femtosecond spin angular momentum transfer
to the lattice, takes place.\cite{Sta07} 

Although the theoretical research in this field has been most 
intense, understanding the microscopic mechanisms of UFD and angular-momentum relaxation still remains an open 
problem.\cite{Tur13} Over the past years, two different theoretical approaches have
attracted particular attention. One of them is electron-phonon spin-flip 
scattering, in which the lattice is assumed to be a perfect sink for 
angular momentum.\cite{Koo05,Ste09,Ste10,Fae11,Mue13} 
The other one is spin-polarized electron diffusion, which does not invoke any
angular momentum dissipation channel, but rather a spin-dependent 
superdiffusive electron transport from the laser-excited film to the 
substrate.\cite{Bat10,Mel11,Bat12,Sch13} 
In this context it is quite remarkable that none of these theories 
happens to bear a clear relation with the fundamentals of itinerant 
magnetism, which are tightly anchored to strong electron correlations and 
to the resulting high stability of the local $3d$ magnetic 
moments.\cite{mag-temp} Only recently the potential importance of localized spins and their 
fluctuations has been suggested.\cite{Sch13-loc,Ill13} 
It is therefore most challenging to establish the links between 
equilibrium and non-equilibrium theories of itinerant magnetism. 
The purpose of this Letter is to develop a many-body theory of UFD,
which takes into account the electronic correlations responsible for
local moment formation and magnetic order, and to analyze its 
physical consequences rigorously by performing exact time propagations.

Several distinct features are expected to be central to the physics of 
the laser-excited electrons in FM metals:
(i)~the single-particle hybridizations responsible for electron 
delocalization, bonding and metallic behavior, 
(ii)~the dominant Coulomb interactions among the $3d$ electrons, 
which introduce correlations, stabilize the local magnetic moments and, 
together with the single-particle contributions, define the magnetic order, 
(iii)~the SO interactions, which couple the spin and orbital 
degrees of freedom, and 
(iv)~the interaction with the external laser field, which triggers
the initial electronic excitation.
We therefore propose the $pd$-band model given by 
\begin{equation}
\label{eq:modelham}
\hat H = \hat H_0 + \hat H_{C} + \hat H_\text{SO} + \hat H_{E}(t)  \, ,
\end{equation}
where
\begin{equation}\label{eq:tightbinding}
\hat H_0 = \sum_{i \alpha \sigma} \varepsilon_\alpha \hat n_{i \alpha \sigma} + 
           \sum_{ij} \sum_{\alpha \beta \sigma} \, t_{ij}^{\alpha \beta} \, 
           \hat c_{i\alpha\sigma}^\dagger \hat c_{j\beta\sigma}
\end{equation}
describes the band structure of the relevant $3d$ and $4p$ valence
electrons that are responsible for the magnetic properties in TMs 
and for the relevant optical absorption. 
The operator $\hat c_{i\alpha\sigma}^\dagger$ ($\hat c_{i\alpha\sigma}$) 
creates (annihilates) a spin-$\sigma$ electron at atom $i$ in the 
orbital $\alpha$, which has well-defined radial and 
orbital quantum numbers $nlm$. 
$\hat n_{i \alpha \sigma} = \hat c_{i\alpha\sigma}^\dagger 
\hat c_{i\alpha\sigma}$ counts the corresponding occupations. 
The energy of the orbital $\alpha$ is denoted by $\varepsilon_\alpha$
and the interatomic hopping integrals by $t_{ij}^{\alpha \beta}$.
The second term in Eq.~(\ref{eq:modelham}) designates the dominant 
intra-atomic Coulomb interaction among the $3d$ electrons
\begin{equation}
\label{eq:interaction}
\hat H_{C} = \dfrac{U}{2} \sum_i \hat n_i^d ( \hat n_i^d - 1) 
- J \sum_i \hat{\vec s}_i^{\,\, d} \cdot \hat{\vec s}_i^{\,\, d} \, ,
\end{equation}
where $U$ stands for the average $d$-electron direct Coulomb integral 
and $J$ for the exchange integral.\cite{uchida,gari}
The operators $\hat n_i^d$ and $\hat{\vec s}_i^{\,\, d}$ refer, respectively, 
to the $d$-electron number and total spin at atom $i$.
The spin-orbit interactions are given by the third term 
\begin{equation}\label{eq:soc}
\hat H_\text{SO} = \xi\, \sum_{i} \sum_{\alpha \beta \in 3d} \sum_{\sigma \sigma'} 
                  ( \vec l \cdot \vec s )_{\alpha \sigma , \beta \sigma'} \,
                  \hat c_{i\alpha\sigma}^\dagger \, \hat c_{i\beta\sigma'} \, ,
\end{equation}
where $( \vec l \cdot \vec s)_{\alpha \sigma , \beta \sigma'}$ 
stands for the intra-atomic matrix elements of $\vec l \cdot 
\vec s$ and $\xi$ is the SOC constant. For simplicity, 
$4p$ electrons are here ignored. 
Finally, the last term 
\begin{equation}\label{eq:dipole-approx}
\hat H_E(t) %= e \hat{\vec r} \cdot \vec E(t) 
            = e \vec E(t) \cdot \sum_{i\alpha\beta\sigma} \, 
              \bra{\alpha} \hat{\vec r} \ket{\beta} \, 
              \hat c_{i\alpha\sigma}^\dagger \, \hat c_{i\beta\sigma}
\end{equation}
describes the interaction with the external laser field $\vec E(t)$ 
in the intra-atomic dipole approximation ($e > 0$ is the electron charge).  
The usual atomic selection rules for the position operator $\hat{\vec r}$ 
imply that only $dp$ and $pd$ transitions enter the sum. 

At this point it is useful to recall the fundamental conservation laws 
underlying the model, which are the same as in the first-principles
Hamiltonian. The non-relativistic terms $\hat H_0$, $\hat H_C$ and 
$\hat H_E$ conserve the total spin $\vec S = \sum_{i\alpha} \vec s_{i\alpha}$, 
since $[\hat H_0, \vec S] = [\hat H_C , \vec S] = [\hat H_E , \vec S] = 0$. 
The spin conservation is broken only by 
$\hat H_\text{SO} \propto 
\sum_{i} [(l_{i+} s_{i-} + l_{i-} s_{i+})/2 + l_{iz} s_{iz}]$, 
which involves intra-atomic angular-momentum transfer between $\vec s_i$ 
and $\vec l_i$. Still, the total angular momentum 
$\vec j_i = \vec l_i + \vec s_i$ is locally conserved in the SO
transitions, since $[\hat H_\text{SO}, \vec s_i + \vec l_i] = 0$ 
for any atom $i$. Intra-atomic Coulomb interactions are also invariant upon
rotations and thus preserve $\vec s_i$, $\vec l_i$ and $\vec j_i$.
However, the interatomic hybridizations, though total-spin conserving, do not
preserve the orbital angular momentum 
$\vec L = \sum_{i\alpha} \vec l_{i\alpha}$, since the hopping integrals
$t_{ij}^{\alpha \beta}$ connect orbitals having different $m$ at different atoms 
($[\hat H_0, \vec L] \not= 0$ and $[\hat H_0, \vec l_i] \not= 0$).\cite{sym-lin}
We shall see that these simple arguments allow us to understand a number of
qualitative aspects of the magnetization dynamics, its dependence on the 
relevant physical parameters, and the main mechanism behind it.

Despite the simplicity and transparency of the proposed $pd$-band model, 
an analytical or straightforward numerical solution of its dynamics 
is out of reach at present. As in the equilibrium case, the main 
difficulties stem from the the Coulomb interaction $\hat H_C$ and the
resulting many-body behavior. One could in principle resort to time-dependent 
mean-field approximations. However, these are known to introduce 
artificial symmetry breakings, which spoil the fundamental spin rotational
invariance, thus casting potentially serious doubts on the resulting 
magnetization dynamics. In order to derive rigorous conclusions, 
we have therefore decided to consider a simplified small-cluster 
version of the model and to obtain an exact numerical solution of 
the ground state, excitations and time propagation. Similar approaches 
have been most successful in the context of equilibrium properties of 
narrow-band systems.\cite{Vic85,dagotto,prlclus} 
In addition, our results show that the physics of the magnetization dynamics can be explained by short-range effects, 
which justifies the small-cluster approach \textit{a posteriori}.

The parameters used for the calculations correspond approximately to Ni 
and have been specified as follows. The nearest-neighbor (NN) 
Slater-Koster hopping integrals are taken from band structure 
calculations.\cite{Papaconstantopoulos} The $dp$ promotion energy $\Delta\varepsilon_{pd}= 1$~eV 
yields a dominant $3d$ band occupation in the ground state. The direct Coulomb integral 
$U = 4.5$~eV and the exchange integral $J = 0.8$~eV lead to FM order with nearly saturated $3d$ moments at low 
temperatures.\cite{Vic85} Finally, the SOC constant $\xi$ is varied in the range $|\xi| \le 100$~meV,
which includes the values found in $3d$ TMs.\cite{foot:xi}
For the numerical calculations we reduce the degeneracy of the bands, 
by considering only three $3d$ orbitals per atom (having $m = -1$, $0$ and $1$) and one $4p$ orbital per atom ($m=0$).
The fcc (111) monolayer is modeled by an equilateral triangle 
having $N_e = 4$ valence electrons. Since the 
average occupation of the $d$ orbitals is below half-band filling ($\langle \hat n_{i\alpha}^d \rangle \simeq N_e / 3 
N_a = 0.44$) we set $\xi < 0$ in order that $\vec s_i$ and $\vec l_i$ align parallel to each other.\cite{foot:signxi}
 
A first test on the validity of the model and parameter choice is provided 
by the ground-state results, which match qualitatively the magnetic 
properties of Ni. From exact Lanczos diagonalizations we obtain that the 
ground state is FM with an off-plane easy magnetization 
axis and a magnetic anisotropy energy $\Delta E = 2.8$~meV per atom. 
The local spin momenta $2 s_{iz} = 1.32 \hbar$ are almost saturated, 
whereas the local orbital momenta $l_{iz} = 0.09 \hbar$ are quenched
to a large extent. These values should be compared, for example, with 
$2 s_{iz} = 0.62 \hbar$ and $l_{iz} = 0.07 \hbar$ obtained in experiment.\cite{Sta10} 

Starting from the FM ground state we determine the exact time evolution numerically by performing short-time iterative 
Lanczos propagations.\cite{Tan07} The actual dynamics is triggered by the femtosecond laser pulse 
$\vec E (t) = \vec \varepsilon \cdot E_0 \cos (\omega t) \exp(-{t^2}/{\tau_p^2})$,
whose polarization vector $\vec \varepsilon$ is along a NN 
bond within the (111) plane. The laser wave length $\lambda = 800$~nm 
corresponds to the photon energy $\hbar \omega = 1.55$~eV used in several 
experiments.\cite{Hoh97,Rhi03,Cin06,Sta07,Car08,Sta10} 
The pulse is centered at $t = 0$ and has a duration of $\tau_p = 5$~fs. 
Its amplitude $E_0$ is such that the electronic system absorbs $0.2$ 
photons per atom on average, which corresponds to a fluence $F \simeq 40$~mJ/cm$^2$.\cite{foot:param}

\begin{figure}[t]
\centering
\includegraphics[width=8.5cm,keepaspectratio=true]{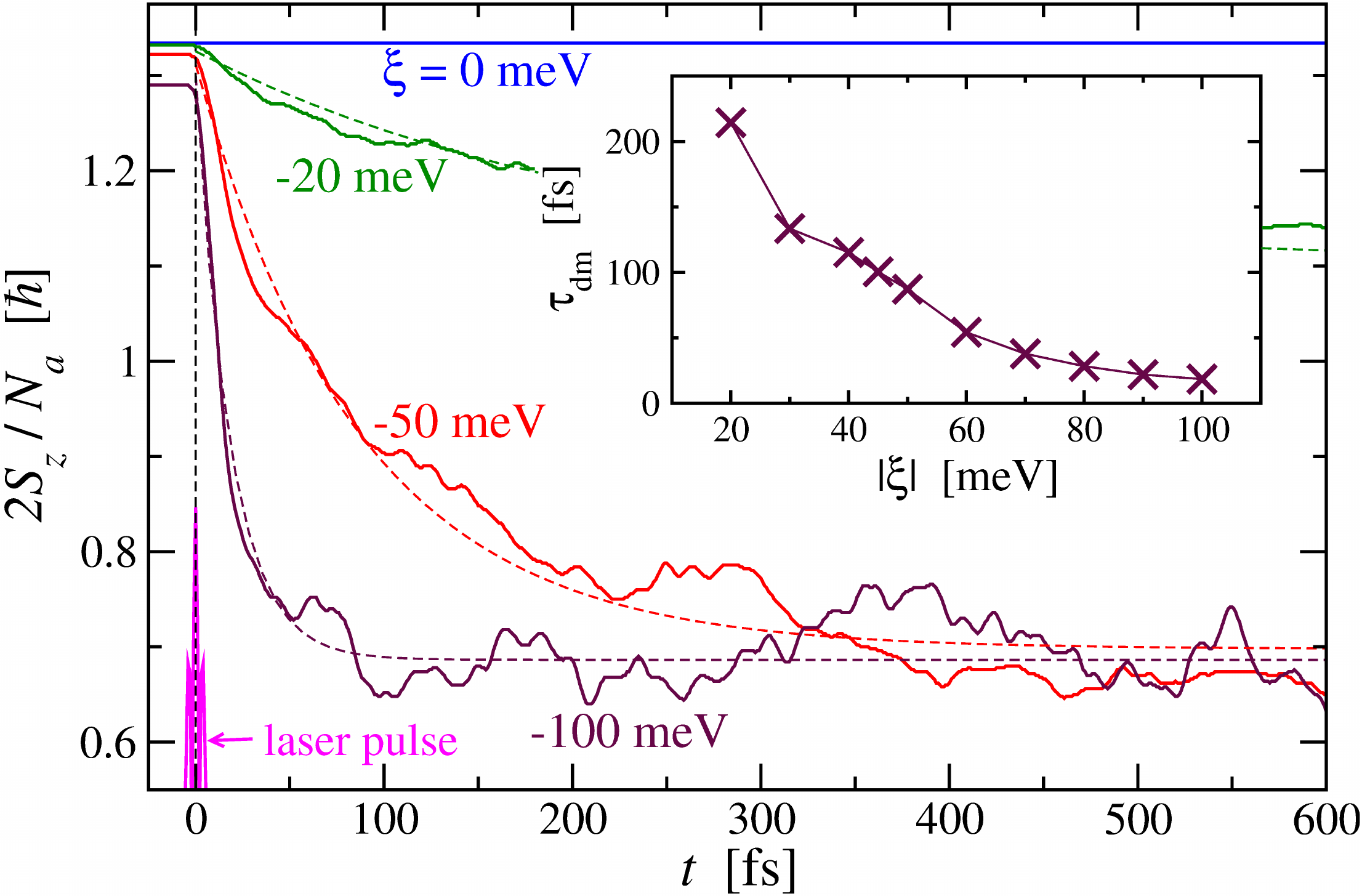} %width=8.663cm
% figure1.eps: 0x0 pixel, 300dpi, 0.00x0.00 cm, bb=(atend)
\caption{(Color online) 
%Exact calculated 
Time dependence of the spin magnetization $2 S_z$ following a $5$~fs laser-pulse excitation for an equilateral triangle 
having $N_e = 4$ valence electrons and different SOC strengths $\xi$. The inset shows the demagnetization time 
$\tau_\text{dm}$ as a function of $\xi$.}
\label{fig:soc}
\end{figure}

In Fig.~\ref{fig:soc} results are given for the magnetization dynamics. 
For realistic values of $|\xi| = 50$--$100$~meV the spin angular momentum 
per atom relaxes irreversibly from $2 S_z / N_a = 1.32 \hbar$ to around $0.64 \hbar$ 
within the first hundreds of femtoseconds following the laser excitation. 
This demonstrates the ultrafast demagnetization effect in agreement with 
experiment.\cite{Sta07} Fig.~\ref{fig:soc} also reveals the central role played by the SOC.
For $\xi = 0$ the total spin is unaffected by the optical excitation, 
as expected. Moreover, as the SOC is turned on one observes that 
the degree of demagnetization $\Delta S_z = S_z(0) - S_z (\infty)$ 
increases and that the  demagnetization time $\tau_\text{dm}$ decreases.
Finally, for larger $|\xi| \ge 50$~meV the SOC strength affects only the 
time scale. In order to quantify $\tau_\text{dm}$, we have fitted the
exact time dependence $S_z(t)$ with an exponential function of the form 
$\tilde S_z (t) = \Delta S_z \exp(-t/\tau_\text{dm}) + S_z(0) - \Delta S_z$, which is shown by the dashed curves in 
Fig.~\ref{fig:soc}. The obtained $\tau_\text{dm}$, shown in the inset, behaves approximately as 
$\tau_\text{dm} \approx 6 {\hbar}/{|\xi|}$. This is consistent with 
the time-energy uncertainty relation and confirms that the
SOC, which represents the smallest energy scale, controls the time scale
of the relaxation process.

\begin{figure}[t]
\centering
\includegraphics[width=8.5cm,keepaspectratio=true]{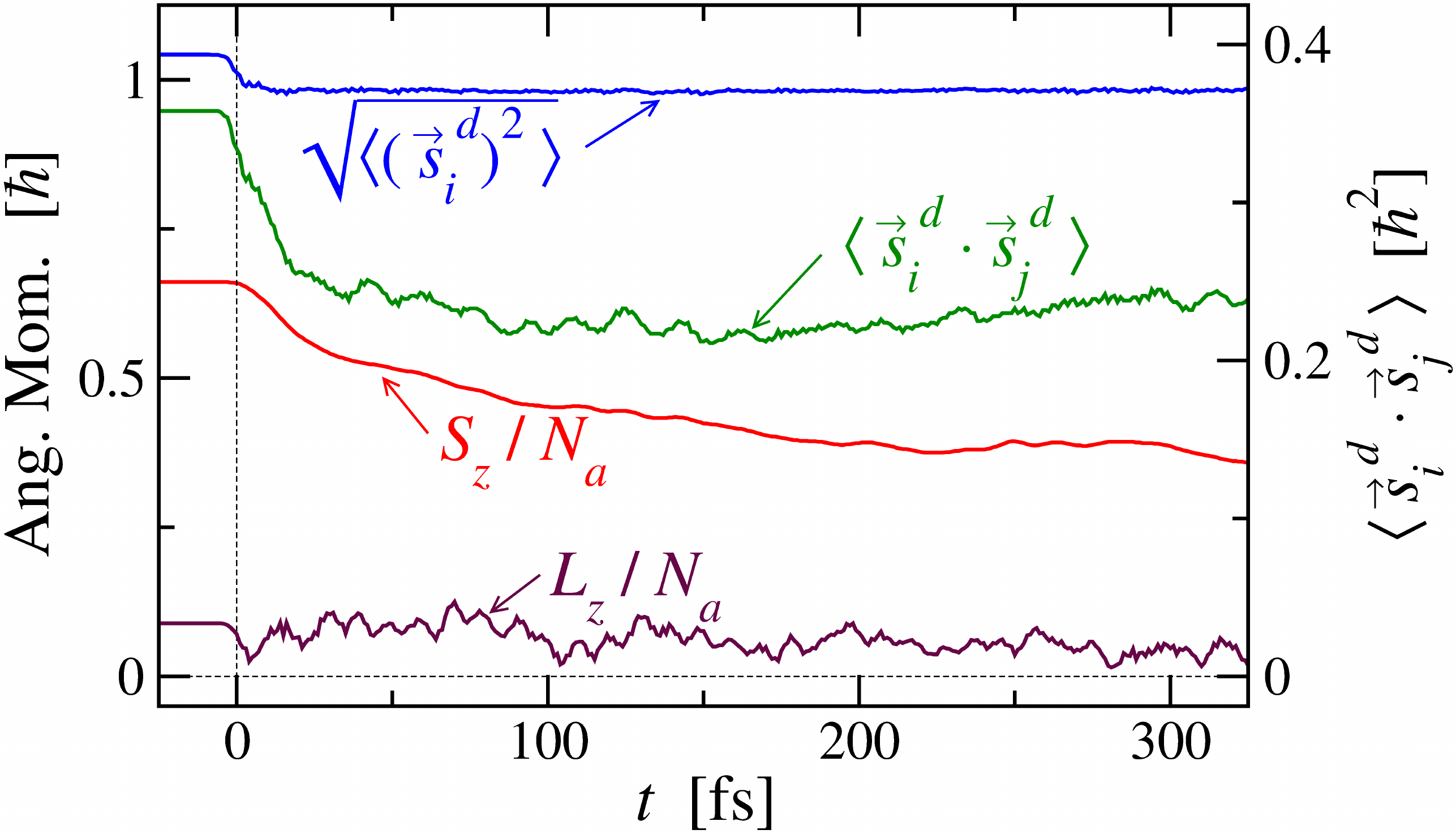}
% figure2.eps: 0x0 pixel, 300dpi, 0.00x0.00 cm, bb=(atend)
\caption{(Color online)
Time dependence of the average $d$-electron local spin moment 
$\langle (\vec{s}_{i}^{\,\,d})^2 \rangle^{1/2}$, NN spin-spin 
correlation function $\langle \vec s_i^{\,\, d} \cdot \vec s_j^{\,\, d} \rangle$, and off-plane spin and orbital 
angular momentum $S_z$ and $L_z$.}
 \label{fig:how}
\end{figure}

The time dependences of the local $d$-electron spin moments 
$\mu_i^d = \sqrt{\langle (\vec s_i^{\,\, d})^2 \rangle }$ and of the
NN spin correlation functions $\gamma_{ij} = \langle \vec s_i^{\,\, d} \cdot \vec s_j^{\,\, d} \rangle$
shown in Fig.~\ref{fig:how} for $\xi = -50$~meV provide a much clearer picture of 
how the UFD actually occurs. One observes that $\mu_i$ changes very
slightly from $1.04 \hbar$ to $0.98 \hbar$ while the laser pulse is on ($\tau_p = 5$~fs). This is the result of a small 
laser-induced $dp$ charge transfer, which involves majority $d$ electrons, and which causes the number of $d$-electrons 
per atom to decrease from $n_d \simeq 1.25$ to $n_d \simeq 1.17$.
Otherwise, $\mu_i^d$ is essentially time independent. 
The remarkable stability of the local $d$ moments shows,
as in thermal equilibrium above the Curie temperature $T_C$, that
the laser-induced ultrafast breakdown of FM order is not the 
consequence of a significant loss of local spin polarization.\cite{mag-temp} 
Instead, the UFD follows to a large extent from the quantum fluctuations 
of the {\it orientations} of the local magnetic moments $\vec s_i^{\,\, d}$, 
whose magnitude remains almost constant throughout the process.  
Indeed, as shown in Fig.~\ref{fig:how}, the average NN spin-correlation 
function $\gamma_{ij}$ decreases very rapidly as a function of time, approaching its long-time 
limit already $50$-$80$~fs after the laser absorption, and remaining 
approximately constant in the following. The breakdown of 
the NN correlations is actually much more rapid than the decrease of the
total spin polarization $S_z$ (see Fig.~\ref{fig:how}). The same trend holds for other values of $\xi$, for example 
$\xi = -100$~meV, where the demagnetization time is much shorter.
This shows that the quantum fluctuations of $\vec s_i^{\,\, d}$ 
dominate only the first stages of the spin dynamics. A slower
tilting of $\vec S$ off the $z$ axis follows, during which 
the transversal components $S_x$ and $S_y$ always remain zero.

In order to analyze the mechanisms of angular-momentum transfer behind the UFD process, we turn our attention to the 
dynamics of the orbital moment $L_z$ and contrast it with the dynamics of $S_z$ (see Fig.~\ref{fig:how}).
In the ground state, before the laser excitation, $L_z / N_a \simeq 0.09 \hbar$ is 
quenched to a large extent in comparison with the Hund-rule atomic value 
$L_z^{\rm at} = 1 \hbar$, as expected in TMs. After the excitation $L_z / N_a$ shows rapid oscillations between 
$0.02 \hbar$ and $0.12 \hbar$, always remaining parallel to $S_z$. This 
oscillatory behavior is numerically stable and perfectly reproducible. The total angular momentum $J_z = L_z + S_z$ is 
obviously not a constant of motion. This is in agreement with time-resolved XMCD 
experiments showing that $L_z$ is not a reservoir for the decreasing 
$S_z$.\cite{Sta07,Sta10,Boe10} 
The fact that $L_z$ remains quenched for all times is the result of 
the interatomic hybridizations responsible for the electronic motion 
in the lattice and for the band formation. Formally, one could say that 
the operator $\hat H_0$ preserves $\vec S$ but not $\vec L$ 
(i.e., $[\hat H_0 , {\vec S}] = 0$ while $[\hat H_0 , {\vec L}] \neq 0$) 
since the potential generated by the ions is not rotationally invariant.\cite{Zha08} 
But physically it is important to realize that the characteristic time 
$\tau_q$ required to quench the orbital moment of an electron in a 
lattice is extremely short, of the order of 
$\tau_q \sim \hbar/W_d \simeq 0.1$~fs where $W_d\simeq 6$~eV stands 
for the $d$-band width.\cite{landau} 
The combination of a local $\vec J$-conserving transfer of angular momentum 
from $\vec S$ to $\vec L$, due to SO interactions, and a very fast 
dynamical quenching of $\vec L$, due to electronic hopping, explains the 
spin-to-lattice angular-momentum transfer observed in experiment.
In the framework of the present model, the dynamical quenching of $L$
would manifest itself only as a global rotation of the rigid lattice,
since electron-phonon coupling (EPC) has been neglected. Including EPC
in the model would open an additional channel for $L$-quenching, through
which angular momentum would be transferred to the lattice vibrations.

In order to verify the validity of our conclusions we have repeated 
the calculations of the dynamics by reducing artificially the hopping 
integrals $t_{ij}^{\alpha \beta}$ and thus approaching the atomic zero-band-width limit.
As shown in the Supplemental Material, one observes that as soon as 
$t_{ij}^{\alpha \beta}$ becomes comparable or smaller than the SOC constant $\xi$ the 
rapid quenching of $L_z$ is replaced by oscillations of both $L_z$ 
and $S_z$, with a period of about $70$~fs, during which $J_z$ is approximately 
conserved.\cite{supp-mat}

\begin{figure}[t]
\centering
\includegraphics[width=8.5cm,keepaspectratio=true]{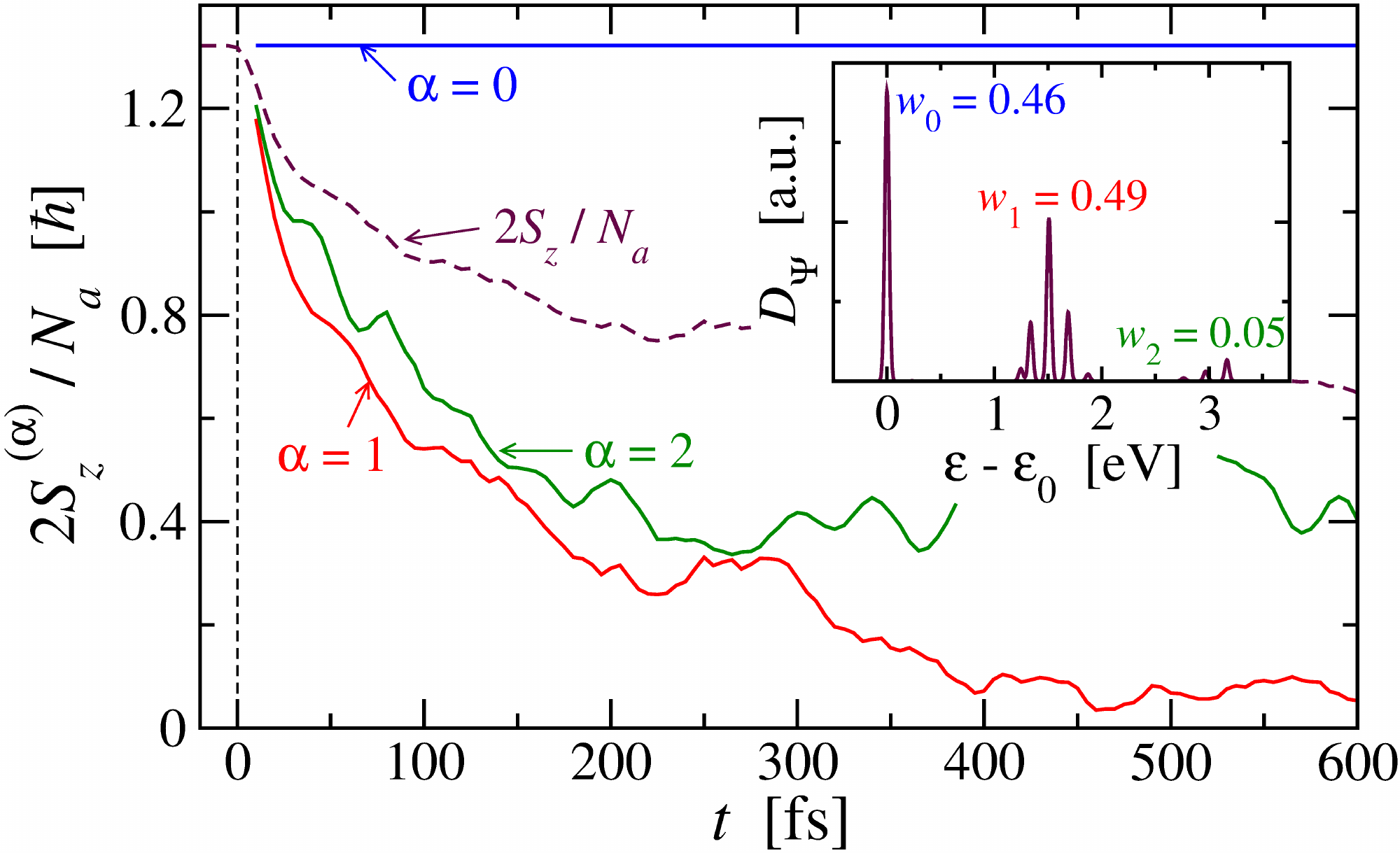}
% figure3.eps: 0x0 pixel, 300dpi, 0.00x0.00 cm, bb=(atend)
\caption{(Color online) 
Time dependence of the spin magnetization $2 S_{z}^{(\alpha)}$ in the 
excited-state manifolds corresponding to $\alpha = 0$, $1$ and $2$ photon absorptions ($\hbar \omega = 1.55$~eV). 
The average $S_z = \sum_\alpha w_\alpha S_{z}^{(\alpha)}$ is given 
by the dashed curve. The inset shows the spectral distribution $D_{\Psi}(\varepsilon)$ of $\Psi(t)$ after 
the laser-pulse passage ($t \geq 3 \tau_p = 15$~fs). The weights $w_\alpha$ of the 
different spectral parts of $\Psi$ are indicated.}
 \label{fig:spectral}
\end{figure}

A complementary perspective to the UFD process is obtained by performing a 
spectral analysis of the laser-excited many-body state $\Psi (t)$ and of the
time dependence of $S_z$. For this purpose, we considered $\Psi (t)$ 
immediately after the excitation (at $t = 3\tau_p$) and expanded it in 
the stationary states $|\psi_k\rangle$ of the field-free Hamiltonian 
$\hat H = \hat H_0 + \hat H_{C} + \hat H_\text{SO}$ having energy $\varepsilon_k$. 
The obtained spectral distribution 
$D_{\Psi} (\varepsilon) = \sum_{k} \delta(\varepsilon - \varepsilon_k) 
| \skp{\psi_k}{\Psi} |^2$ is shown in the inset 
of Fig.~\ref{fig:spectral}. Notice that $D_{\Psi} (\epsilon)$ is independent 
of $t$ after the pulse passage, since so is $\hat H$ 
[see Eqs.~(\ref{eq:modelham})--(\ref{eq:dipole-approx})]. 
Three main groups or manifolds of nearby peaks can be recognized, 
which are located at the excitation energies 
$\Delta\varepsilon \simeq \alpha\hbar\omega$ and 
which correspond to the absorption of $\alpha = 0$, $1$ and $2$ photons. 
The spin magnetization $S_{z}^{(\alpha)}$ originating from these 
manifolds is shown in Fig.~\ref{fig:spectral} as a function of $t$.  
One observes that immediately after excitation all $\alpha$ have the same 
saturated magnetization. This simply reflects the spin conservation upon
optical dipole transitions. The ground state, being a pure stationary 
state, preserves $\langle \hat S_z \rangle$ and yields a time-independent 
$S_{z}^{(0)}$. Relaxation can only stem from the excitations. 
Indeed, in the excited manifolds $S_{z}^{(\alpha)}$ decreases dramatically 
to nearly zero, particularly in the most relevant $\alpha = 1$ manifold. 
This demonstrates the remarkably high efficiency of laser-induced UFD 
in the excited states. The residual average magnetization $2 S_z(t\to\infty)$, 
which persists after the spin-lattice relaxation process (see Fig.~\ref{fig:soc}) is essentially the consequence of the 
finite overlap between the many-body state $\Psi(t)$ after the laser absorption and the ground state.

In sum, the laser-induced magnetization dynamics of ferromagnetic TMs has been 
studied in the framework of an electronic model. For the first time a 
solution of the time-dependent many-body problem has been achieved, which 
is complete from the perspectives of electron correlations, 
spin-orbit interactions and essential symmetries. The results have demonstrated that the femtosecond demagnetization 
can be explained in terms of a three-step mechanism: (i)~The laser pulse creates electron-hole pairs. This opens the 
way for (ii) the SOC yielding local angular-momentum transfer from $\vec s_i$ to $\vec l_i$ with a characteristic time 
scale of $\hbar / |\xi| \approx 10$~fs. However, angular momentum is not accumulated in $\vec l_i$, since (iii)~$\vec 
L$ is quenched by the motion of electrons in the lattice. This takes place on a much shorter time scale of only $\hbar 
/ W_d \lesssim 1$~fs. Extensions of this work by improving the model Hamiltonian are certainly 
desirable and necessary for a more realistic description of specific magnetic 
materials. In particular the coexistence of different relaxation channels (electronic relaxation,
electron-phonon spin-flip scattering, spin diffusion, etc.) deserves to be explored. In any case, the
simplicity of the electronic processes identified in this work and the fundamental character of the proposed model 
suggest that the rigorously derived concepts should be universally applicable.

%
%\begin{acknowledgments}
W.~T.\ acknowledges the support provided by a fellowship of the 
Otto-Braun Foundation. Computer resources were supplied by the 
IT Service Center of the University of Kassel and by the CSC of the
University of Frankfurt. 
%\end{acknowledgments}

%

%

%

\begin{thebibliography}{99}
\bibitem{mag-temp}
The local magnetic moments of $3d$ TMs are known to
remain stable even well above the Curie temperature $T_C$.
In a previous density-functional study we have shown that the Ni moments are extremely stable even at very high 
levels of thermalized electronic excitation. See Ref.~\onlinecite{Toe13}.
\bibitem{sym-lin}
Only in the presence of a very high point-group symmetry some 
component of $\vec L$ may be conserved, for example, $L_{z}$ in 
a linear chain along the $z$ axis.
\bibitem{foot:signxi}
The sign of $\xi$ can be changed by performing the electron-hole
transformation $\hat h_{i\alpha\sigma} = c_{i\alpha\sigma}^\dagger$, which 
does not affect the Coulomb interaction and only changes the sign of the 
hopping integrals. Thus, band fillings above one half can be 
simulated with less than one fermion per orbital. We have 
explicitly checked that changing the sign of $\xi$ does not affect the
time dependence of the discussed observables in any significant way,
except for the change of the relative orientation between 
$\vec s_i$ and $\vec l_i$.
\bibitem{foot:param}
It has been explicitly checked that reasonable changes in the model parameters do not affect our physical conclusions.
\end{thebibliography}
\end{document}